\documentclass[aps,pra,amsmath,amsfonts,amssymb,twocolumn,showpacs]{revtex4}
\usepackage[latin1]{inputenc}
\usepackage{graphicx}
\usepackage[english]{babel}
\usepackage[usenames,dvipsnames]{color}

\usepackage{amsfonts}
\usepackage{amsmath}
\usepackage{amssymb}

\newcommand{\beq}{\begin{equation}}
\newcommand{\eeq}{\end{equation}}
\newcommand{\nn}{\nonumber}
\newcommand{\ket}[1]{|#1\rangle}
\newcommand{\bra}[1]{\langle #1|}

\makeatletter
\@ifundefined{textcolor}{}
{%
 \definecolor{BLACK}{gray}{0}
 \definecolor{WHITE}{gray}{1}
 \definecolor{RED}{rgb}{1,0,0}
 \definecolor{GREEN}{rgb}{0,1,0}
 \definecolor{BLUE}{rgb}{0,0,1}
 \definecolor{CYAN}{cmyk}{1,0,0,0}
 \definecolor{MAGENTA}{cmyk}{0,1,0,0}
 \definecolor{YELLOW}{cmyk}{0,0,1,0}
 }

\begin{document}

\title{Non-Markovianity induced by a single-photon packet in a one-dimensional waveguide}

\author{M. F. Z. Arruda
}

\author{D. Valente
}
\email{daniel@fisica.ufmt.br}

\author{T. Werlang
}
\email{thiago\_werlang@fisica.ufmt.br}

\affiliation{
Instituto de F\'isica, Universidade Federal de Mato Grosso, CEP 78060-900, Cuiab\'a, MT, Brazil}

\begin{abstract}
The concept of non-Markovianity (NM) in quantum dynamics is still an open debate. 
Understanding how to generate and measure NM in specific models may aid in this quest.
In quantum optics, an engineered electromagnetic environment coupled to a single atom can induce NM.
The most common scenario of structured electromagnetic environment is an optical cavity, composed by a pair of mirrors.
Here, we show how to generate and measure NM on a two-level system coupled to a one-dimensional waveguide
with no mirrors required.
The origin of the non-Markovian behavior lies in the initial state of the field, prepared as a single-photon packet.
We analyze how NM depends on two experimentally controllable parameters, namely, the linewidth of the packet and its central frequency.
We relate the presence of NM to a $\pi$-phase shift between incoming and emitted fields.
We also show how the two output channels of the waveguide provide distinct signatures of NM, both experimentally accessible.
\end{abstract}

\pacs{03.65.Yz, 03.67.-a}

\maketitle

\section{Introduction}

The Markov approximation has been widely used to describe the dynamics of a quantum system coupled to its environment \cite{breuer}. 
In this approach, the quantum dynamics can be represented by a completely positive and trace-preserving (CPTP) dynamical map with a generator in the Lindblad form - a quantum markovian master equation \cite{lindblad}. 
However, the Markov approximation can be violated when the coupling between the system and its environment is strong or when there is not a clear separation between the typical timescales associated with the system and the environment. 
Such a scenario has been reached experimentally, for example, in the contexts of quantum biology \cite{fleming07} and condensed matter physics \cite{imamoglu08}, where the non-Markovian effects become relevant. 

Although the concept of non-Markovianity is well-established in the classical case, the definition of non-Markovianity in the context of quantum open systems is still subject of debate in the scientific community. 
For example, the definition proposed in Ref. \cite{plenio10} is based on the notion of non-Markovianity for a classical stochastic processes, namely, a quantum evolution is non-Markovian if it cannot be described by a {\it divisible} completely positive trace preserving (CPTP) map. 
On the other hand, the non-Markovianity has also been characterized by a temporary back-flow of information from the environment to the system \cite{breuer09,breuer12}. 
Other studies suggest that non-Markovianity can also be measured by the rate of change of the volume of accessible states of an open system \cite{mauro13}, the amount of classical information extracted by the environment \cite{felipe14}, the appearance of negative decoherence rates \cite{hall14}, and the smallest amount of isotropic noise that must be added to a quantum dynamics in order to describe it through a memoryless master equation \cite{wolf08}.
Besides its importance from a fundamental perspective, the non-Markovianity can also be useful in applications involving quantum communication \cite{bylicka14}, quantum metrology \cite{chin12}, quantum correlation generation \cite{huelga12}, just to name a few. 

The need to understand and to exploit non-Markovianity creates an interest in generating and manipulating non-Markovian quantum evolutions.
For example, Bi-Heng Liu {\it et al.} \cite{liu11} reported an all-optical experiment where non-Markovian aspects of the dynamics of a qubit - represented by the polarization degree of freedom of a photon - can be controlled by manipulating the initial state of its environment - represented by the frequency degrees of freedom. Another way to control the non-Markovianity of the dynamics of a qubit coupled to its environment is to manipulate the interaction between the qubit and an additional auxiliary qubit \cite{werlang13}, as demonstrated experimentally in a nuclear magnetic resonance setup \cite{souza13}. In the context of cavity quantum electrodynamics, the dynamics of a two-level atom may display non-Markovian effects due to the confinement of the electromagnetic field inside a cavity. This happens because the photon emitted by the atom can be reflected by the mirrors of the cavity and then reabsorbed by the atom. Recently, T. Tufarelli and collaborators \cite{tufarelli14} showed that the dynamics of an atom coupled to a one-dimensional half-cavity - a semi-infinite waveguide with a perfect mirror at one end - can also provide a non-Markovian behavior. In this case, the presence of the mirror is also responsible for the origin of non-Markovianity. 

Here, we show that non-Markovianity in the two-level system (TLS) dynamics is induced by a single-photon packet in a one-dimensional waveguide.
From a fundamental perspective, there is novelty in the physical origin of non-Markovianity in our system.
It is generated by a single-photon pulse, requiring no mirrors at the boundaries of the TLS.
We also provide the means to access the non-Markovianity of the TLS by making measures only on the field.
From the applications perspective, our approach allows for control of non-Markovianity by adjusting the input state of the field, such as its central frequency and linewidth, requiring no change in the system parameters.

This paper is organized as follows. 
In Sec. \ref{sectionmodel}, we introduce the model and solve for the equations of motion.
Both the TLS and the field are studied.
In Sec. \ref{nmmeasure}, we explore the measures of non-Markovianity that are suitable for our model.
In Sec. \ref{results}, we present the signatures of non-Markovianity as they appear both in the TLS (A) and in the field (B) dynamics, both qualitatively and quantitatively.


\section{Model}
\label{sectionmodel}

\begin{figure}[!htb]
\centering
\includegraphics[width=\linewidth]{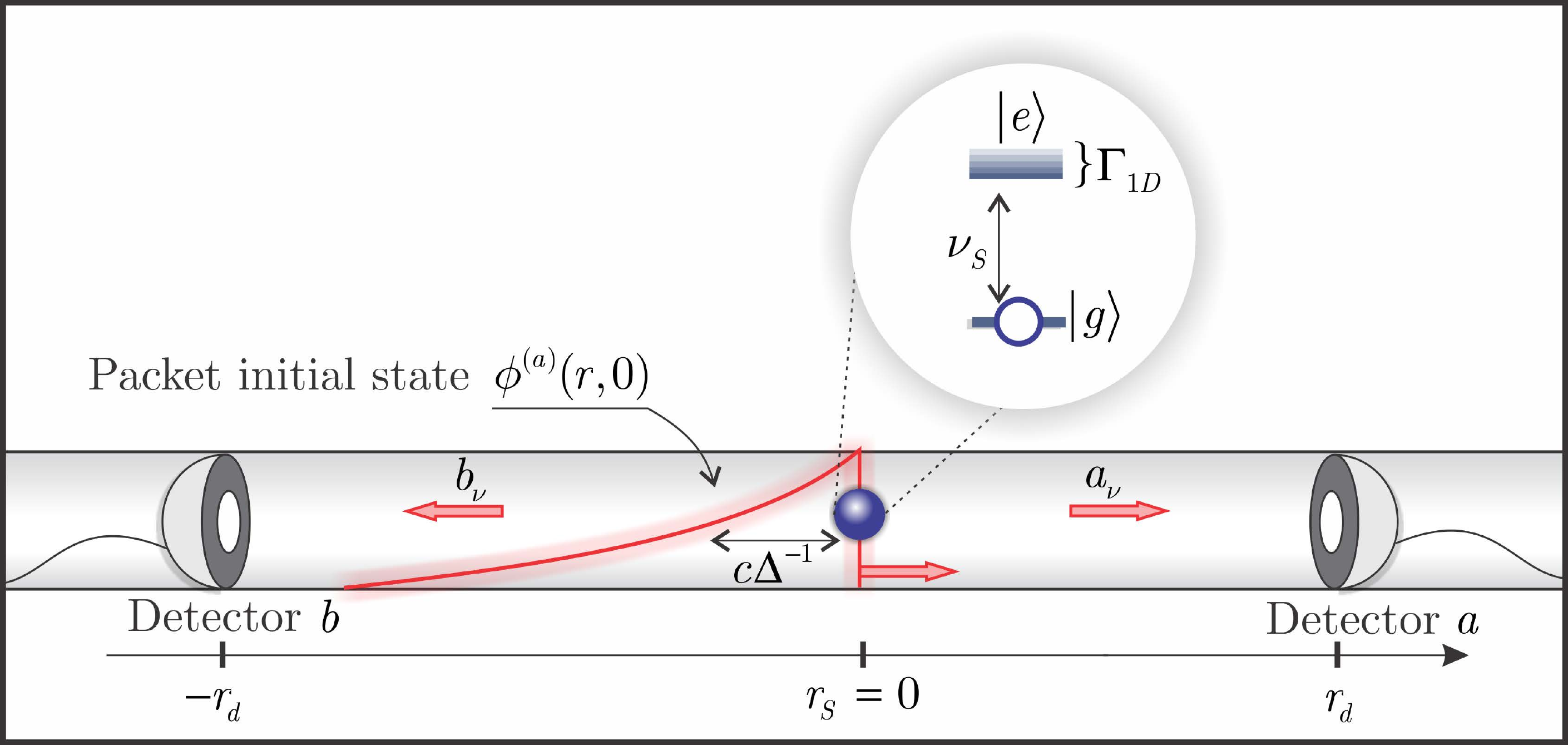}
\caption{(Color online) One-dimensional waveguide coupled to a TLS placed at $r_S=0$. 
The ground and excited states of the TLS are denoted respectively by $\ket{g}$ and $\ket{e}$. 
$\nu_{S}$ is the transition frequency and $\Gamma_{\mathrm{1D}}$ is the spontaneous emission rate of the TLS.
$a_\nu\ (b_\nu)$ denotes the forward (backward) propagating modes of the electromagnetic environment with momentum $\vec{k}_\nu=+|\vec{k}_\nu| \ (\vec{k}_\nu=-|\vec{k}_\nu|)$.
The TLS, initially in the ground state, scatters a single-photon packet described by the initial state $\phi^{(a)}(r,0)$. 
$\Delta^{-1}$ is the typical time duration of the pulse.
The forward (backward) propagating packet after the scattering is measured by detector $a(b)$ located at position $r_d(-r_d)$.} 
\label{model}
\end{figure}

The model we consider (see Fig. \ref{model}) consists of a one-dimensional (1D) infinite waveguide coupled to a two-level system (TLS) \cite{vuckovic1,jmg,japa,tese}, whose ground and excited states are denoted respectively by 
$\ket{g}$ and $\ket{e}$.
The Hamiltonian of the TLS is $H_{S}=\hbar\nu_{S}\sigma_{+}\sigma_{-}$, where $\sigma_{+}=\ket{e}\bra{g}=\sigma_{-}^{\dagger}$ and $\nu_{S}$ is the transition frequency.
The TLS is coupled to a 1D electromagnetic environment composed by a continuum of frequency modes $\nu$. 
Those modes are separated into two channels, $a$ and $b$.  
Channel $a\ (b)$ describes the forward (backward) propagating modes $a_\nu\ (b_\nu)$ with momentum $\vec{k}_\nu=+|\vec{k}_\nu| \ (\vec{k}_\nu=-|\vec{k}_\nu|)$.
The momentum is determined by the dispersion relation $\nu=c|\vec{k}_\nu|=ck_\nu$.
The Hamiltonian of the field is $H_{\mathrm{field}} = \hbar\sum^{\infty}_{\nu=0}\nu \left( a_{\nu}^{\dagger}a_{\nu} + b_{\nu}^{\dagger}b_{\nu} \right)$ and the system-environment interaction Hamiltonian in the rotating-wave approximation (RWA) is
\beq
H_{\mathrm{int}}
=-i\hbar\sum^{\infty}_{\nu=0}g_{\nu}\left[\sigma_{+}\left(a_{\nu}e^{ik_{\nu}r_S}+b_{\nu}e^{-ik_{\nu}r_S}\right) - h.c.\right], 
\label{hint}
\eeq
where $r_S$ is the TLS position, $g_{\nu}$ is the coupling strength between the TLS and the mode with frequency $\nu$, and $h.c.$ is the Hermitian conjugate.
The RWA is valid in the weak coupling regime, that is, $g_{\nu}\ll\nu_S$. 
In the following, we assume the reference frame where $r_S=0$.
The total Hamiltonian is
\beq
H = H_{S} + H_{\mathrm{field}} + H_{\mathrm{int}}.
\label{htotal}
\eeq

Since the number of excitations is conserved, the quantum state of the total system is written in the subspace of zero and one excitation,
\begin{eqnarray}
\ket{\psi(t)} &=& c_{0}(t)\ket{g,0} + \psi_{S}(t)\ket{e,0} + \sum^{\infty}_{\nu=0}\phi^{(a)}_{\nu}(t)a^{\dagger}_{\nu}\ket{g,0} \nonumber \\
&+& \sum^{\infty}_{\nu=0}\phi^{(b)}_{\nu}(t)b^{\dagger}_{\nu}\ket{g,0},
\label{ansatz}
\end{eqnarray}
where $\ket{0}$ is the vacuum state, $\psi_{S}(t)$ is the excited state probability amplitude, $\phi^{(a),(b)}_{\nu}(t)$ is the probability amplitude of a photon with frequency $\nu$ in the channel \textit{a} (\textit{b}), and $c_{0}(t)$ is the probability amplitude corresponding to no excitation. 
The dynamics of the total system is obtained by solving the Schr\"odinger equation, $i\hbar\partial_{t}\ket{\psi(t)} = H\ket{\psi(t)}$, that leads to the differential equations
\begin{eqnarray}
\dot{\psi}_{S}(t) &=& -i\nu_{S}\psi_{S}(t) - \sum_{\nu}g_{\nu}\left(\phi_{\nu}^{(a)}(t) + \phi_{\nu}^{(b)}(t)\right) \label{ppsi} \\
\dot{\phi}_{\nu}^{(a),(b)}(t) &=& -i\nu\phi_{\nu}^{(a),(b)}(t) + g_{\nu}\psi_{S}(t), \label{pphi}\\
\dot{c_{0}}(t) &=& 0.\label{pc0}
\end{eqnarray}
The last equation implies that $c_{0}(t)$ is time independent.

We investigate the case in which the TLS is initially in the ground state, $\psi_S(0)=0$. 
We also assume that the initial state of the field, in the real-space representation $\phi^{(a),(b)}(r,t) \equiv \sum_{\nu}\phi^{(a),(b)}_{\nu}(t)e^{\pm i k_{\nu} r}$ \cite{tese}, is given by
\begin{eqnarray}
\phi^{(a)}(r,0)&=& N\ \Theta(-r) \ e^{\left(\frac{\Delta}{2}+i\nu_L\right)\frac{r}{c}},\nn\\
\phi^{(b)}(r,0)&=&0,
\label{initialfield}
\end{eqnarray} 
where $\nu_L$ is the central frequency of the photon, $\Delta$ is the packet linewidth and $N$ is a normalization constant. 
The typical time duration of the pulse is $\Delta^{-1}$ (see Fig. \ref{model}). 
This initial state can be generated, for example, by the spontaneous emission of an auxiliary TLS, initially in the excited state 
\cite{sand,ck,wallraff}.


\subsection{Two-level System Dynamics}

The reduced dynamics of the TLS can be obtained in the Wigner-Weisskopf approximation \cite{breuer}. 
In this approach, we assume the continuum limit $\sum_{\nu}\rightarrow\int d\nu \rho_{\mathrm{1D}}$, where 
$\rho_{\mathrm{1D}}$ is the density of modes, and $g_{\nu} \approx g_{S}$.
By substituting the solution of Eq.(\ref{pphi}) into Eq.(\ref{ppsi}), the dynamics of the excited state probability amplitude
$\psi_S(t)$ becomes
\begin{eqnarray}\label{dPSIdynamics}
\dot{\psi}_{S}(t) &=& -\left( \frac{\Gamma_{\mathrm{1D}}}{2}+i\nu_{S}\right)\psi_{S}(t)  \\
&-& \sqrt{\frac{\Gamma_{\mathrm{\mathrm{1D}}}}{4\pi\rho_{\mathrm{1D}}}}\left(\phi^{(a)}(-ct,0) + \phi^{(b)}(ct,0)\right)\nonumber,
\end{eqnarray}
where $\Gamma_{\mathrm{1D}}=4\pi g_{S}^{2}\rho_{\mathrm{1D}}$ is the spontaneous emission rate of the TLS in the 
1D waveguide. 
Using the initial conditions from Eqs.(\ref{initialfield}), with $N=\sqrt{2\pi \rho_{1D}\Delta}$, and $\psi_S(0)=0$, the solution of Eq.(\ref{dPSIdynamics}) can be written as
\beq 
\psi_{S}(t) = 
-\sqrt{\frac{\Gamma_{\mathrm{1D}}\Delta}{2}} 
e^{-\left( \frac{\Gamma_{\mathrm{1D}}}{2} + i\nu_{S} \right)t}
\left[
\frac{ e^{\left(\frac{\Gamma_{\mathrm{1D}}-\Delta}{2}-i\delta_{L}\right)t}-1}{\frac{\Gamma_{\mathrm{1D}}-\Delta}{2}-i\delta_{L}}\right],
\label{psis}
\eeq
where $\delta_{L}=\nu_{L}-\nu_{S}$ is the detuning. 
Note that, in the monochromatic limit $\Delta\ll\Gamma_{\mathrm{1D}}$, 
the excited state population tends to $|\psi_S(t)|^2 \leq 2\Delta/\Gamma_{\mathrm{1D}}$.
The behavior of the field in this limit is discussed in the next subsection.

The reduced density operator of the system, $\rho_{S}(t)=\mbox{Tr}_{\mathrm{field}}\left(\ket{\psi(t)}\bra{\psi(t)}\right)$, in the $\left\{\ket{e},\ket{g}\right\}$ basis, is 
\beq
\rho_{S}(t)=
	\begin{pmatrix}
	|\psi_{S}(t)|^{2} & c_{0}^{*}\psi_{S}(t) \\
	c_{0}\psi_{S}^{*}(t) & 1- |\psi_{S}(t)|^{2}
	\end{pmatrix}.
\eeq
The reduced dynamics is given by the Master Equation \cite{breuer}
\begin{eqnarray}\label{meq}
\frac{d}{dt}\rho_{S}(t) &=& -\frac{i}{2}S(t)[\sigma_{+}\sigma_{-},\rho_{S}(t)]  \\
&+& \Gamma(t)\left(\sigma_{-}\rho_{S}(t)\sigma_{+}-\frac{1}{2}\left\{\sigma_{+}\sigma_{-},\rho_{S}(t)\right\}\right)\nonumber,
\end{eqnarray}
where
\beq
\Gamma(t) =  -2\ \mbox{Re}\left(\frac{\dot{\psi}_{S}(t)}{\psi_{S}(t)}\right) 
\label{Gammat}
\eeq
represents the decay rate of the system, being $\mbox{Re}(.)$ the real part, and
\beq
S(t) = -2\ \mbox{Im}\left(\frac{\dot{\psi}_{S}(t)}{\psi_{S}(t)}\right)
\label{St}
\eeq
represents the Lamb-shift, being $\mbox{Im}(.)$ the imaginary part. 
Both quantities depend on time.


\subsection{Field Dynamics in the Real Space Representation}

The dynamics of the field is studied in the real-space representation, as defined in the previous section.
From the solution of Eq.(\ref{pphi}) and applying the Wigner-Weisskopf approximation we find that
\begin{eqnarray}
\phi^{(a)}(r,t) &=& \phi^{(a)}(r-ct,0) \nonumber \\
&+&\sqrt{\pi \Gamma_{\mathrm{1D}} \rho_{\mathrm{1D} }} \Theta(r)\Theta\left(t-\frac{r}{c}\right) \psi_{S}\left(t-\frac{r}{c}\right)
\nn\\
\label{phia}
\end{eqnarray}
and
\beq
\phi^{(b)}(r,t)=
\sqrt{\pi \Gamma_{\mathrm{1D}} \rho_{\mathrm{1D}}} 
\Theta(-r)\Theta\left(t+\frac{r}{c}\right) \psi_{S}\left(t+\frac{r}{c}\right).
\label{phib}
\eeq
The first term in Eq(\ref{phia}), namely $\phi^{(a)}(r-ct,0)$, describes the free propagation of the initial wave packet. 
The second one describes  the amplitude of a photon emitted by the TLS into channel $a$. 
The Heaviside theta functions guarantee that the emitted photon in channel $a$ is detected at $r>0$, after the photon arrived, $t > r/c$, at the detector at position $r$. 
Note that, for vanishing TLS-waveguide coupling $\Gamma_{\mathrm{1D}}\rightarrow 0$, the outgoing field is exactly the input field propagating in the direction defined by channel $a$ at speed $c$, without changing the packet shape. 
For the counter-propagating packet, $\phi^{(b)}(r,t)$, only the amplitude of the emitted photon is present, since the initial photon state contains no amplitude in channel $b$, i.e., $\phi^{(b)}(r+ct,0)=0$.
The Heaviside theta functions in Eq.(\ref{phib}) guarantee that a photon emitted into channel $b$ is detected at $r<0$, after it has arrived, $t>|r|/c$, at the detector at position $-|r|$. 
Eqs.(\ref{phia}) and (\ref{phib}) recover a well-known effect of light-matter interaction in one-dimensional environments, namely, total light reflection \cite {domokos,alexia,vuckovic,japa}. 
This is due to a destructive interference between emitted and incoming fields \cite{tese,kojima,Rabi1D,daniel}. 
It happens at resonance, $\delta_L = 0$, and in the quasi-monochromatic regime, 
$\Delta \lll \Gamma_{\mathrm{1D}}$, for which 
\footnote{ 
From  Eq.(\ref{psis}) at $\delta_L = 0$ and $\Delta \ll \Gamma_{\mathrm{1D}} $, 
$\psi_S(t) \approx  -  \sqrt{2 \Delta/\Gamma_{\mathrm{1D}} } e^{-\Delta t/2} e^{-i\nu_L t}$, 
valid for $t > \Gamma_{\mathrm{1D}}^{-1}$.
So, 
$\sqrt{\pi \Gamma_{\mathrm{1D}} \rho_{\mathrm{1D}}}\  \psi_S(t-r/c) 
\approx  -
N e^{-\Delta(t-r/c)/2} e^{-i\nu_L (t-r/c) } = - \phi^{(a)}(r-ct,0)$, where 
$N = \sqrt{2\pi \Delta \rho_{\mathrm{1D}} }$.
} 
\beq
\phi^{(a)}(r>0,t) \Big|_{\{ \delta_L = 0,\ \Delta \lll \Gamma_{\mathrm{1D}} \} } \approx 0.
\eeq
In the ideal monochromatic regime, no light will propagate through channel $a$ after having interacted with the TLS, being totally reflected into channel $b$. 
Note that the origin of such destructive interference is the $\pi-$phase shift between the incoming and the emitted packets. 
The field emitted into channel $b$ also carries this $\pi$-phase shift with respect to the incoming packet, given that $\phi^{(b)}(r<0,t>|r|/c) \propto \psi_S(t-|r|/c) \propto -\phi^{(a)}(r-ct,0)$.


\section{Measure of non-Markovianity}
\label{nmmeasure}

In recent years different measures have been proposed in order to characterize and quantify the non-Markovian aspects of the dynamics of an open quantum system \cite{plenio10,breuer09,breuer12,mauro13,felipe14,hall14,bylicka14}. 
When the dynamics of the system is described by a time-local master equation
\beq
\frac{d\rho}{dt}=\mathcal{L}(t)\rho(t),
\eeq
where
\begin{eqnarray}\label{generator}
\mathcal{L}(t)&=&-\frac{i}{\hbar}[H(t),\rho(t)]\\
&+&\sum_i\gamma_i(t)\left[L_i(t)\rho L_i^\dagger(t)-\frac{1}{2}\left\{L_i^\dagger(t)L_i(t),\rho\right\}\right]\nonumber,
\end{eqnarray}
with
\beq
\mbox{Tr}\left[L_i(t)\right]=0 \quad\mbox{e}\quad \mbox{Tr}\left[L_i^\dagger(t)L_j(t)\right]=\delta_{ij},
\eeq
is a time-dependent generator of the dynamics in canonical form \cite{breuer12,hall14}, all measures agree that the quantum dynamics is Markovian if the time-dependent decay rates are always non-negative, that is, $\gamma_i(t)\geq0$. 
In this case, the generator (\ref{generator}) is in Lindblad form \cite{lindblad} for each fixed $t\geq0$. 
On the other hand, if the decay rates assume negative values, the dynamics of the system is {\it indivisible}, that is, the quantum dynamics cannot be described by a sequence of infinitesimal CPTP quantum evolutions \cite{plenio10,breuer12}. 
According to the measure proposed by Rivas, Huelga and Plenio (RHP) \cite{plenio10}, a quantum dynamics is non-Markovian if it corresponds to an indivisible CPTP quantum dynamics. 

Since the dynamics of the system of interest, as given by Eq. (\ref{meq}), is described by a time-local master equation, the temporary appearance of a negative decay rate will be used as a witness for non-Markovianity.
In other words, the TLS dynamics is non-Markovian if, and only if, $\Gamma(t)<0$ for some $t>0$. 
According to Ref. \cite{plenio10}, the amount of non-Markovianity is given by
\beq
\mathcal{N}=\int_{0}^{\infty}f(t)dt,
\label{nonm}
\eeq
with $f(t)\equiv\max\left\{0,-\Gamma(t)\right\}\geq0$, where $\Gamma(t)$ is the decay rate given by Eq. (\ref{Gammat}). 
For the quantum dynamics described by Eq. (\ref{meq}), the negativity of the decay rate can also be used as a signature of the non-Markovian behavior according to the measures proposed by Lorenzo {\it et al.} \cite{mauro13}, Hall {\it et al.} \cite{hall14}, and Wolf {\it et al.} \cite{wolf08}.   Regarding the measure proposed by Breuer {\it et al.} \cite{breuer09}, the non-Markovianity is not necessarily guaranteed when $\Gamma(t)<0$.

In our case, the non-Markovianity can be characterized by the dynamics of the excited state population $|\psi_S(t)|^2$. 
To show this, note that Eq.(\ref{Gammat}) can be rewritten as
\beq
\Gamma(t)=-\frac{1}{|\psi_S(t)|^2}\frac{d|\psi_S(t)|^2}{dt}.
\label{gammapop}
\eeq
Thus, the decay rate becomes negative, $\Gamma(t)<0$, when there is an increase in the excited state population, $d|\psi_S(t)|^2/dt>0$.


\section{Results and Discussions}
\label{results}

In this section we show the signatures of non-Markovianity evidenced both in the TLS dynamics (A) and
in the field dynamics (B). 
In subsection A, we first present the existence of oscillations in the population of the excited state.
We also show that these oscillations imply negative values for $\Gamma(t)$, revealing non-Markovianity.
The physical origin and meaning of such oscillations is explained.
Finally, we quantify the amount of non-Markovianity in terms of the detuning $\delta_L$ and the packet linewidth $\Delta$.
In subsection B, we explore the means to directly access both $\Gamma(t)$ and the TLS excited state population  experimentally, by making measurements only on the field.


\subsection{Signatures of Non-Markovianity in the Two-Level System Dynamics}

\begin{figure}[!htb]
\centering
\includegraphics[width=\linewidth]{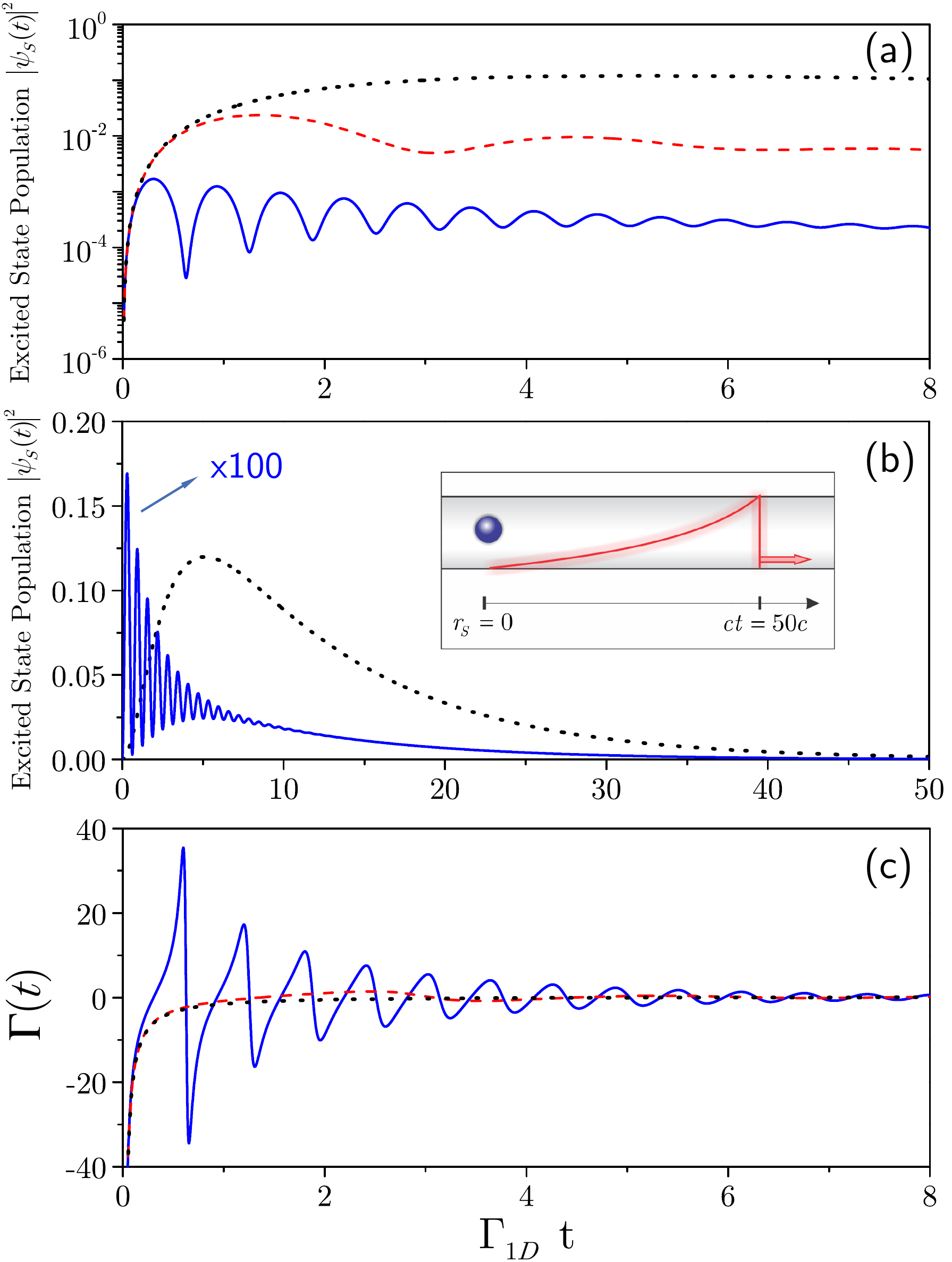}
\caption{
(Color online) 
(a) Excited state population of the TLS, $|\psi_S(t)|^2$, as a function of time for $\Delta=0.1$ and $\delta_L=0$ (black dotted line), $\delta_L=2$ (red dashed line) and $\delta_L=10$ (blue solid line). 
The population exhibits an oscillatory behavior as the detuning increases.
(b) Dynamics of the excited state population $|\psi_S(t)|^2$ for a time interval that is extended until $\Gamma_{\mathrm{1D}} t = 50$, keeping $\Delta=0.1$. 
The black dotted curve corresponds to $\delta_L = 0$ and the blue solid line corresponds to $\delta_L = 10$.
The curve associated with $\delta_L = 10$ has been multiplied by a factor $100$ so it fits the same vertical axis scale.
The inset shows that for $ct=50c$ the photon packet has passed completely through the TLS.
Thus, after $\Gamma_{\mathrm{1D}} t = 50$, the TLS reaches the steady state, where the excited state population becomes zero.
(c) Time-dependent decay rate $\Gamma(t)$ as a function of time for $\Delta=0.1$ and the same values of $\delta_L$ used in (a). 
The negative values of the decay rate, $\Gamma(t)<0$, indicate that the TLS dynamics is non-Markovian. 
Note that $\Gamma(t)<0$ when there is an increase of the excited state population, $d|\psi_S(t)|^2/dt>0$.
We set $\Gamma_{\mathrm{1D}} = 1$ as our frequency unit in all plots.}
\label{gammaepsi}
\end{figure}

The dynamics of the excited state population of the TLS, $|\psi_S(t)|^2$, is plotted in Fig.\ref{gammaepsi}-(a).
The analytical expression is obtained from Eq.(\ref{psis}).
We set $\Gamma_{\mathrm{1D}} = 1$ as our frequency unit in all plots.
The packet linewidth is kept at $\Delta = 0.1$. 
The curves correspond to $\delta_L = 0$ (black dotted line), 
$\delta_L = 2$ (red dashed line) and $\delta_L = 10$ (blue solid line).
The frequency of the population oscillation increases with the detuning, so as the number of peaks.

We emphasize that the TLS is not in steady state for times $\Gamma_{\mathrm{1D}} t \sim 8$.
In Fig.\ref{gammaepsi}-(b) we plot the same quantity as in figure (a),
but for a time interval that is extended until $\Gamma_{\mathrm{1D}} t = 50$, 
keeping $\Delta = 0.1$. 
The two curves correspond to $\delta_L = 0$ (black dotted) and $\delta_L = 10$ (blue solid line). 
Note that the latter has been multiplied by a factor $100$ so it fits the same vertical axis scale.
In both cases, it is clear that the TLS is left in the ground state after the photon packet amplitude at the TLS position vanishes.
This is illustrated in the inset.
The time of interaction between the photon packet and the TLS scales as $\Delta^{-1}$.

Fig.\ref{gammaepsi}-(c) shows $\Gamma(t)$ for the same time interval as in figure (a), i.e., until 
$\Gamma_{\mathrm{1D}} t = 8$.
Again, the same detunings have been chosen, namely,
 $\delta_L = 0$ (black dotted line), $\delta_L = 2$ (red dashed line) and $\delta_L = 10$ (blue solid line).
As discussed in section III, at any given time $t$ for which the excited state population is  increasing, $d|\psi_S(t)|^2/dt > 0$, 
the decay rate becomes negative, $\Gamma(t)<0$.
It is particularly noticeable at higher detunings, e.g., at $\delta_L = 10$.
In Sec.
III, we stated that $\Gamma(t) < 0$ implies the presence of non-Markovianity for the master equation of our model.
Hence, population oscillation, or more precisely population increase, is a signature of non-Markovianity in the
TLS dynamics.

The physical origin of the oscillations in the TLS populations shall be explained.
Firstly, it must be made clear that these are not Rabi oscillations \cite{Rabi}, neither classical \cite{Rabi1D} nor quantum \cite{QuantumRabi}.
It is not classical because the input is a quantum pulse, containing a single photon.
It is not quantum Rabi oscillations either, since the photon emitted by the TLS should be otherwise reabsorbed after having been reflected by the mirrors of a cavity. 
In the present scenario, the emitted photon is never reabsorbed. 
The excited state amplitude dynamics, $\partial_t \psi_S(t)$, exclusively depends on the component of the photon packet that precedes the TLS, $\phi^{(a)}(-ct,0)$, not on $\phi^{(a)}(r>0,t)$ neither $\phi^{(b)}(r<0,t)$, which propagate outwards the waveguide.
The remarkable feature of those oscillations is their dependence on the detuning.
The number of peaks is proportional to $\delta_L/\Gamma_{\mathrm{1D}}$.
This fact indicates that this effect is phase dependent, suggesting that it can be written in terms of an interference between two amplitudes, $A_1$ and $A_2$.
Interference between those two amplitudes is quantified by $\mathcal{I} = |A_1+A_2|^2-(|A_1|^2+|A_2|^2)$.
In Sec.\ref{sectionmodel} we analyzed the interference effect in the field, between incoming and emitted photon amplitudes.
On the contrary, we now seek amplitudes describing different states of the TLS itself.
For that purpose, Eq.(\ref{dPSIdynamics}) can be recasted as 
$\partial_t \tilde{\psi}_S = A_1+A_2 $,  where we define 
$A_1\equiv-(\Gamma_{\mathrm{1D}}/2)\  \tilde{\psi}_S(t)$ and 
$A_2 \equiv -\sqrt{(\Gamma\Delta)/2}\ \tilde{\phi}^{(a)}(t) \exp(-i\delta_L t)$.
Note that $\phi^{(b)}(ct,0)=0$ due to the chosen initial condition.
A rotating frame has been applyied, $\psi_S(t) = \tilde{\psi}_S(t) \exp(-i\nu_S t)$ and 
$\tilde{\phi}^{(a)}(t) \equiv \exp\left(-\Delta t/2  \right)$.
We rewrite the increase of the excited state population, $\partial_t |\tilde{\psi}_S|^2 > 0$, as $\partial_t |A_1|^2 >0$.
It follows that
\begin{eqnarray}
\partial_t |A_1|^2 &=& -\frac{\Gamma_{\mathrm{1D}}}{2} \left(  2|A_1|^2  + A^*_2 A_1 + A_2 A^*_1 \right) \nn\\
 &=&-\frac{\Gamma_{\mathrm{1D}}}{2} \left(  2|A_1|^2 + \mathcal{I}  \right).
\label{interf}
\end{eqnarray}
The interference term $\mathcal{I}$ in Eq.(\ref{interf}) can be interpreted as a quantum interference between the amplitudes of the TLS to have already been excited ($A_1$) and to have not been excited yet ($A_2$) hence being in the ground state. 
In the latter case, it is the incoming field that carries the excitation.
If no interference takes place, $\mathcal{I} = 0$, Eq.(\ref{interf}) reduces to spontaneous emission behavior, 
$\partial_t |A_1|^2=-\Gamma_{\mathrm{1D}}\ |A_1|^2$. 
No increase of population is expected in this case. 
Only destructive interference, $\mathcal{I}<0$, can induce increase of the excited state population.
We now analyze the dynamics of the interference term, $\mathcal{I}(t)$.
Close to resonance for a quasi-monochromatic packet, $\delta_L \ll \Delta \ll \Gamma_{\mathrm{1D}}$, 
the dynamics of interference is $\mathcal{I}(t) \approx - 2\ |A_1(t)| |A_2(t)|\ \mbox{Re}[ \exp(-i\delta_L t) ]$. 
Therefore, in that regime, interference is periodically destructive with frequency $\delta_L$, as qualitatively expected.
Destructive interference becomes even more evident precisely at resonance, 
$\delta_L = 0$ ($\Delta \lll \Gamma_{\mathrm{1D}}$), for which 
${\psi}_S(t) \propto - \phi^{(a)}(-ct,0)$. 
In that case,  $A_1 = |A_1| = - A_2$ and $\mathcal{I}(t) \approx - 2\ |A_1(t)| |A_2(t)| < 0$.
Such $\pi-$phase shift between these two amplitudes is the same as the one responsible for zero transmission of the light field, $\phi^{(a)}(r>0,t) \approx 0$, as discussed in Sec.\ref{sectionmodel}.
Even in the resonant finite-width packet regime $\Delta \lesssim \Gamma_{\mathrm{1D}}$ there is still a peak in the population curve, as shown in Fig.\ref{gammaepsi}-(b).
In that regime, we have that $A_1=|A_1|$ and $A_2 = - |A_2|$, so destructive interference is maintained. 
However, after a given time $t$, the excited state population amplitude $|A_1(t)|$ becomes bigger than the incoming-packet amplitude $|A_2(t)|$ so the system excitation declines, 
$\partial_t |A_1|^2 = -\Gamma_{\mathrm{1D}}\  |A_1|\ (|A_1|-|A_2|) <0$.
For the large-detuning quasi-monochromatic regime, $\delta_L \gg \Gamma_{\mathrm{1D}} \gg \Delta$, 
we have that $\mathcal{I}(t) \approx - |A_1(t)| |A_2(t)|\ \mbox{Re}[\exp(-i(\delta_L t + \pi))]$. 
Again, interference periodically becomes destructive with frequency $\delta_L$, 
so the peaks in Fig.\ref{gammaepsi}-(a) are spaced by a time interval of nearly $2\pi /\delta_L$.
This last result is valid for both the low and the high detuning regimes.
We have, therefore, evidenced the physical origin of the successive increases in population,
connecting them to $\pi-$phase shifts that appear periodically.

\begin{figure}[!htb]
\centering
\includegraphics[width=\linewidth]{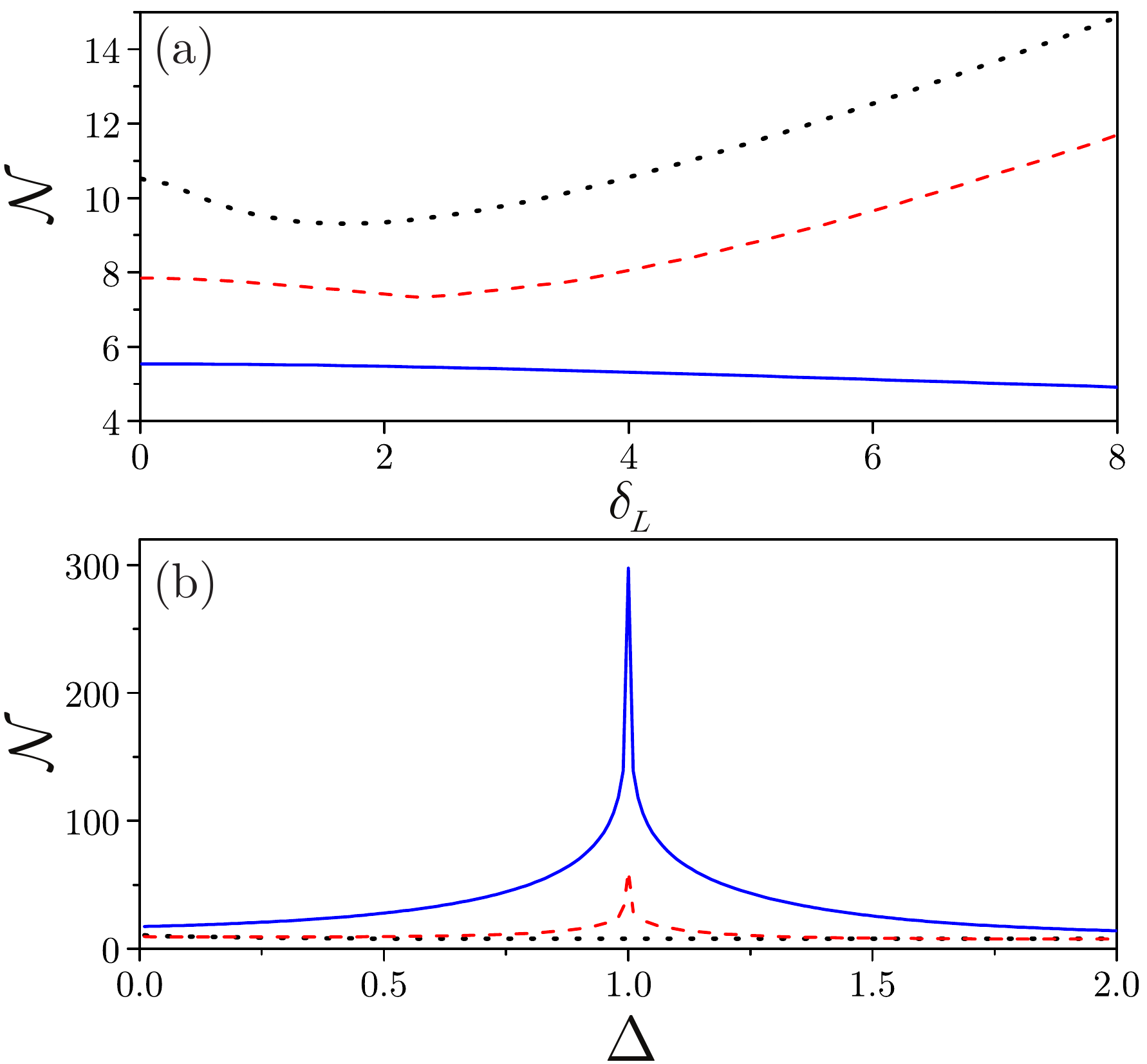}
\caption{(Color online) 
(a) Non-Markovianity $\mathcal{N}$ as a function of detuning $\delta_L$ for $\Delta=0.1$ (black dotted line), $\Delta=2$ (red dashed line), and $\Delta=10$ (blue solid line).
Note that, for $\delta_L\gg\Delta$ the non-Markovianity rises almost linearly.
This is because the excited state population oscillates rapidly under this condition, leading to an increase of the time integral Eq.(\ref{nonm}).
(b) Non-Markovianity $\mathcal{N}$ as a function of linewidth $\Delta$ for $\delta_L=0$ (black dotted line), $\delta_L=2$ (red dashed line), and $\delta_L=10$ (blue solid line).
For $\delta_L\neq 0$, non-Markovianity is maximum at $\Delta=1$.
As we adopt $\Gamma_{\mathrm{1D}} = 1$, the maximum occurs when $\Delta=\Gamma_{\mathrm{1D}}$. In this regime, light-TLS interaction is enhanced, inducing stronger oscillations in the excited state population.}
\label{nmxdelta}
\end{figure}

We now quantify the amount of non-Markovianity $\mathcal{N}$ in terms of the detuning $\delta_L$ and the packet linewidth $\Delta$.
Fig.\ref{nmxdelta}-(a) gives the detuning dependency of the non-Markovianity, $\mathcal{N}(\delta_L)$.
The curves are obtained by the use of Eq.(\ref{nonm}).
Three linewidth values have been chosen, $\Delta = 0.1$ (black dotted line), $\Delta = 2$ (red dashed line) and $\Delta = 10$ (blue solid line).

Three regions can be identified in the plot, corresponding to distinct behaviors.
At very small detunings $\delta_L \ll \Delta$, non-Markovianity $\mathcal{N}$ is nearly constant.
This is due to the tiny deviation from the ideal case of perfect resonance.
The system dynamics is almost unchanged in that case.
At intermediate detunings $\delta_L \sim \Delta$, a decrease of $\mathcal{N}$ becomes apparent.
The amount of time during which there is increase in population gets reduced as detuning grows. 
That is the reason for the reduction of non-Markovianity.
At large detunings $\delta_L \gg \Delta$, non-Markovianity rises almost linearly.
As it has been discussed, large detunings induce fast population oscillations.
The successive peaks in the excited state population yield large $\mathcal{N}$ when integrated in time.
This third region where $\mathcal{N}$ grows is not illustrated in the plot for the largest linewidth, 
$\Delta/\Gamma_{\mathrm{1D}} = 10$.
In such case, the photon packet is so short in time (and space), $\sim \Delta^{-1}$, that no time is left for the population of the system to oscillate during the field passage. However, an even higher detuning will eventually rise non-Markovianity by inducing faster oscillations than the time spent by the packet to go through the TLS.

Fig.\ref{nmxdelta}-(b) shows the linewidth dependency of the non-Markovianity, $\mathcal{N}(\Delta)$,
for the detuning values of $\delta_L = 0$ (black dotted line), $\delta_L = 2$ (red dashed line) and $\delta_L = 10$ (blue solid line).
The most prominent feature is the peak at $\Delta = 1$, at non-zero detunings.
The size of the peak increases with detuning.
It is worth reminding that the unit here is the TLS natural linewidth $\Gamma_{\mathrm{1D}}$.
So, $\Delta = \Gamma_{\mathrm{1D}}$ precisely corresponds to the mode matching condition between the linewidth of the incoming photon and that of the TLS.
Therefore, light-TLS interaction is enhanced, inducing stronger oscillations in the excited state population.
The peaks are asymmetric.
In the large linewidth limit, non-Markovianity vanishes, whereas in the small linewidth limit, non-Markovianity is finite.
Note that $\mathcal{N}$ does not vanish in the 
$\Delta/\Gamma_{\mathrm{1D}}\rightarrow 0$ limit even at resonance, 
because of the unavoidable initial increase in $|\psi_S(t\rightarrow 0^+)|^2$, starting at zero.
Interestingly, the peak is absent for $\delta_L = 0$.
In that case, oscillations in the excited state population are minimized, as previously discussed.
The monotonic decrease of $\mathcal{N}$ with increasing $\Delta$ at zero detuning is reasonable, 
since non-Markovianity is inversely proportional to the excited state population, Eq.(\ref{gammapop}).

Our results are strongly based on the initial condition we have chosen.
In Ref. \cite{tufarelli14}, a different initial condition is addressed on a very similar context, namely, 
the TLS in the excited state and the field in the vacuum state inside a 1D waveguide. 
In that case, non-Markovianity takes place due to a single mirror put close to the atom. 
Non-Markovianity is then shown to vary with respect to the atom-mirror distance. 
In our model, on the other hand, the control on the degree of non-Markovianity comes from the choice of the shape and the frequency of the photon packet.
This allows for dynamical control without the need to change the system parameters, yielding a possible experimental advantage.
We finish this section by calling attention to the importance of a genuine single-photon packet for inducing non-Markovianity.
Usual experiments with 1D waveguides performed at the single-photon level employ very low-excitation laser pulses \cite{lodahl}.
Weak laser pulses can always be modeled by time-dependent Hamiltonians \cite{Rabi}, which in turn induce unitary dynamics.
However, unitary dynamics are Markovian, since negative decay rates do not appear, as shown in Eqs.(\ref{generator}) and (\ref{nonm}).
Therefore, for our system of interest, an authentic single-photon packet in the input field is necessary for generating finite non-Markovianity.


\subsection{Signatures of Non-Markovianity in the Field Dynamics}

So far we have investigated the TLS dynamics. 
From an experimental perspective, accessing the system dynamics can only be done by measuring
the emitted field. 
In 1D electromagnetic environments, however, there can exist interference between the incoming and the emitted fields, as previously mentioned.
So we need to carefully specify what kind of information is encoded in the field propagating towards each of the two detectors  placed outside the waveguide.
The intensity of a photodetection signal is
$I_{a,b}(t) = \epsilon^2 |\phi^{(a),(b)}(r_d,t)|^2$, where $\epsilon$ is the vacuum field constant 
and $r_d$ is the position of the detector.
This relation is in agreement with Glauber's photodetection theory \cite{glauber}.

Firstly, we analyze the signal from the backward-propagating packet (channel $b$), detected at position $r_d < 0$. 
Eq.(\ref{phib}) shows that $I_b(t)$ provides direct access to the excited state population of the TLS,
$I_b(t) \propto |\phi^{(b)}(r_d,t)|^2 = \left| \psi_S \left(t-|r_d|/c\right) \right|^2$.
Therefore, the TLS behavior would be directly measured as shown in Fig.\ref{gammaepsi}-(a).
The blinking of the TLS in the monitor of detector $b$ would indicate the existence of non-Markovianity, 
manifested by successive increases and decreases of population. 

The signal from the forward-propagating packet (channel $a$), however, exhibits richer phenomena.
Eq.(\ref{phia}) evidences the interference between the incoming packet and the one emitted by the TLS.
Sections \ref{sectionmodel} and \ref{results} present detailed analysis on the consequences of that interference.
Being more specific, it has been shown that the $\pi-$phase shift is responsible for both the total reflection of light and the increase of the excited state population, that evidences non-Markovianity.
Hence, destructive interference induces asymmetry between backward and forward propagating fields on the one hand and non-Markovianity on the other hand. 
It is natural, then, to look for a signature of non-Markovianity on the ratio between 
$\phi^{(a)}(|r_d|,t)$ and $\phi^{(b)}(-|r_d|,t)$.
Indeed, it can be shown that
\beq
\Gamma_{\mathrm{1D}}\ \mbox{Re}\left[\frac{\phi^{(a)}(|r_d|,t)}{\phi^{(b)}(-|r_d|,t)}\right] = 
\Gamma \left(t - \frac{|r_d|}{c} \right),
\label{fieldNM}
\eeq
where the term $\Gamma(t)$, on the right side, is defined in Eq.(\ref{Gammat}). Non-Markovianity is exactly quantified by its negativity.
It is worth mentioning that $S(t-|r_d|/c) =-2\nu_S + \mbox{Im}[\phi^{(a)}(|r_d|,t)/\phi^{(b)}(-|r_d|,t)]$, where $S(t)$ is the time-dependent Lamb-shift also defined in Eq.(\ref{St}).
Remarkably, Eq.(\ref{fieldNM}) provides the means to directly obtain the curve shown in Fig.\ref{gammaepsi}-(c) in an experiment,
by exclusively measuring the outgoing electromagnetic fields in the appropriate manner.

\section{Conclusions}

We have shown that a single-photon wave packet generates non-Markovianity in the dynamics of a two-level system coupled to a one-dimensional waveguide.
Both the TLS and the field dynamics have been obtained.
We derived an effective master equation for the TLS open dynamics.
The form of the master equation allowed us to measure non-Markovianity in terms of the negativity of the time-dependent decay rate.
The decay rate is shown to be negative if and only if the time derivative of the excited state population is positive.
Oscillations in the excited state population are explained in terms of periodic $\pi$-phase shifts that appear between incoming and emitted fields.
Non-Markovianity $\mathcal{N}$ is quantified in terms of packet linewidth and detuning.
$\mathcal{N}$ increases almost linearly at large detunings and presents a peak at the mode matching condition $\Delta = \Gamma_{\mathrm{1D}}$.
Both linewidth and detuning are experimentally controllable parameters, yielding a practical advantage.
The signatures of non-Markovianity can be accessed by measuring the light at both ends of the waveguide.
The intensity of the backward propagating field is shown to be proportional to the excited state population of the TLS.
The asymmetry between forward and backward propagating fields reveals non-Markovianity, providing information on the time-dependent decay rate.
We highlight two perspectives offered by this work.
From an experimental point of view, measuring spectra is generally easier than measuring time resolved quantities.
Therefore, relating non-Markovianity of the TLS dynamics to the spectra of the light field in both channels of the waveguide can be of high interest.
From a theoretical point of view, it is relevant to understand the influence of other quantum states of the field on the non-Markovianity of the TLS.
In particular, if the TLS is initially excited and a photon packet is prepared in the waveguide, we could relate stimulated emission to non-Markovianity.
In that case, analyzing Breuer's measure of non-Markovianity would be suitable.
Another interesting problem is to analyze the TLS dynamics after a two-photon packet is prepared inside the waveguide. Temporal correlations between the two photons could generate unexpected NM behaviors.

\begin{acknowledgements}
M.F.Z.A. acknowledges financial support from CAPES (Brazil). 
D.V. acknowledges financial support from CNPq (Brazil) through Grant No. 477612/2013-0.
T.W. acknowledges financial support from CNPq (Brazil) through Grant No. 478682/2013-1. 
\end{acknowledgements}

\end{document}